\documentclass[12pt]{article}
\usepackage{setspace}
\usepackage[margin=1.2in]{geometry}
\usepackage{amsmath}
\usepackage{amsfonts}
\usepackage{multicol}
\usepackage{empheq}
\usepackage{caption}
\usepackage{float}
\usepackage{hyperref}
\usepackage{amsrefs} 
\usepackage{amssymb}
\usepackage{amsthm}
\usepackage{array}
\usepackage{bm}
\usepackage{datetime}
\usepackage{amsmath}
\usepackage{fancyhdr}
\usepackage{fontenc}
\usepackage{ifthen}
\usepackage{accents}
\usepackage{tabularx}

\newcommand{\Schrodinger}{Schr\"odinger }
\newcommand{\D}{{\cal D} }
\begin{document}
\title{\bf Galilean Relativity and the Path Integral Formalism in Quantum Mechanics}
\author{ 
Charles Torre\\
{\sl Department of Physics}
\\ {\sl Utah State University}
\\{\sl Logan, Utah }
\\{\sl USA 84322-4415}
\\
\\
}
\date{}                           

\numberwithin{equation}{section}

\maketitle
 \thispagestyle{empty}
\vskip 0.5truein
\begin{abstract}
Closed systems in Newtonian mechanics obey the principle of Galilean relativity.  However,  the usual Lagrangian for Newtonian mechanics, formed from the difference of kinetic and potential energies, is not invariant under the full group of Galilean transformations.  In quantum mechanics Galilean boosts require a non-trivial transformation rule for the wave function and a concomitant ``projective representation'' of the Galilean symmetry group.  Using Feynman's path integral formalism this latter result can be shown to be equivalent to the non-invariance of the Lagrangian.  Thus, using path integral methods,  the representation of certain symmetry groups in quantum mechanics can be simply understood  in terms of the transformation properties of the classical Lagrangian and conversely.   The main results reported here should be accessible to students and teachers of physics -- particularly classical mechanics, quantum mechanics, and mathematical physics -- at the advanced undergraduate and beginning graduate levels, providing a useful exposition for those wanting to explore topics such as the path integral formalism for quantum mechanics, relativity principles,  Lagrangian mechanics, and representations of symmetries in classical and quantum mechanics.

\end{abstract}

\onehalfspace

\section{Introduction}

Many features of quantum mechanics are made more transparent when viewed from the perspective of Feynman's path integral formalism. See, for example,  references \cite{Feynman1965, Schulman1981, Marinov1980}.  In this article we will show how one can use the path integral to illuminate the implementation of Galilean relativity  in mechanics. 

The group of Galilean transformations -- space-time translations, orthogonal transformations of space, and non-relativistic boosts -- are  space-time symmetries which reflect the Galilean relativity principle in mechanics \cite{Landau1976, Arnold1989}.  For closed Newtonian systems these transformations are symmetries of the equations of motion, mapping solutions to solutions. However,  the usual Lagrangian for Newtonian mechanics, formed from the difference of kinetic and potential energies, is not invariant under the full group of Galilean transformations. In particular, the Lagrangian is not strictly invariant under the subgroup of boosts, but instead acquires a total derivative term.  While it is possible to construct an equivalent Lagrangian for Newtonian mechanics which {\it is} invariant with respect to boosts, one does this at the expense of manifest translational symmetry in the Lagrangian.  One can show that this state of affairs is unavoidable:  there is no Lagrangian for Newtonian mechanics which is strictly invariant under the whole Galilei group 
\cite{Azcarraga1995, Olver2022}.\footnote{By ``Newtonian mechanics'' we mean a system of particles obeying Newton's laws and described by a Lagrangian which is equivalent to the difference of kinetic and potential energies.  If one allows more general equations of motion there {\it are} Galilean-invariant Lagrangians involving higher order derivatives, which can be constructed using the methods of \cite{Olver2022}.}

In non-relativistic quantum mechanics, described by a Hamiltonian which is the sum of kinetic and potential energies, wave functions transform in a non-trivial way with respect to boosts \cite{Wigner1952, Bargmann1954, Hamermesh1960, Levy1963, Brown1999, Landau1977, Merzbacher1998,Ballentine2014}.  This leads to the important result that the Galilei group is ``projectively represented'' -- usually viewed as an intrinsically quantum mechanical possibility.  Indeed, according to \cite{Wigner1952, Bargmann1954, Hamermesh1960, Levy1963, Ballentine2014} the projective representation of the Galilei group is mandatory if one wants a quantum mechanical system to have physically acceptable properties.  See ref.~\cite{Brown1999} for a more comprehensive list of citations to the extensive literature on Galilean symmetry in quantum mechanics. 

It turns out that these two aspects of Galilean symmetry -- the non-existence of a strictly invariant Lagrangian and the projective representation in quantum mechanics -- are ultimately two facets of the same mathematical feature of the Galilei group: its Lie algebra cohomology \cite{Azcarraga1995, Olver2022}.  However, one need not adopt this relatively deep point of view to understand the essence of the connection between the Galilean transformation of the Lagrangian and the representation of the group on wave functions. The purpose of this note is to show how one may understand the  relationship in a relatively straightforward way using Feynman's path integral.   

The main results reported here should be accessible to students and teachers of physics -- particularly classical mechanics, quantum mechanics, and mathematical physics -- at the advanced undergraduate and beginning graduate levels.  Hopefully this article will provide a useful exposition for those wanting to explore topics such as the path integral formalism for quantum mechanics, relativity principles,  Lagrangian mechanics, and representations of symmetries in classical and quantum mechanics.

In the next section  we review the structure of the Galilei group.  In \S 3 we review the transformation properties of the Lagrangian for Newtonian mechanics with respect to the Galilei group.  In \S 4 we review Feynman's path integral formalism and use it  to derive the transformation law of wave functions with respect to boosts. In \S 5 we show how to use a boost-invariant Lagrangian for Newtonian mechanics to derive a non-trivial transformation law of wave functions with respect to space-time translations. In \S 6 we show how the transformation properties of the wave functions under the Galilei group yield the well-known projective representation.   Appendix A demonstrates how Lie algebra cohomology determines the (non-)existence of strictly invariant Lagrangians. Appendix B shows how the same cohomology determines the projective representation. 

\section{The Galilei group}
Any inertial reference frame (IRF) in which Newton's laws apply is a set of events labeled by 4 coordinates $(t, x, y, z) \equiv (t, {\bf r})$.  Here $(x, y, z)$ label events which are simultaneous; the geometry of this 3-dimensional space is Euclidean and  $(x, y, z)$ will be chosen to be Cartesian coordinates. The group of Galilean transformations is generated by combining time translations, space translations,  orthogonal transformations of space, and  ``boosts'', respectively defined by
\begin{align}
&t \longrightarrow t^\prime = t + a, \quad a \in {\bf R}
\label{timetrans}\\
&{\bf r} \longrightarrow{\bf r}^{\, \prime} = {\bf r} + {\bf b}, \quad {\bf b} \in {\bf R}^3
\label{spacetrans}\\
&{\bf r} \longrightarrow {\bf r}^{\, \prime} = { O}\cdot {\bf r}, \quad { O}^{\scriptscriptstyle T} { O} = {\bf 1}
\label{rot}\\
&{\bf r} \longrightarrow {\bf r}^{\, \prime} = {\bf r} + t\,{\bf  u},\quad  {\bf  u} \in {\bf R}^3.
\label{boost}
\end{align}
The primed  labels denote the ``active'' transformation of events.  One can also interpret these formulas from the ``passive'' point of view, in which the events are unchanged but the IRF is  transformed.  From the point of view of two IRFs (primed and un-primed)  eq.~(\ref{timetrans}) has the time of the primed IRF  shifted by $a$ units into the past relative to the un-primed reference frame; in eq.~(\ref{spacetrans}) the primed reference frame its spatial origin displaced by $-{\bf b}$ relative to the un-primed origin; in eq.~(\ref{rot}) the  spatial axes of the primed reference frame are related by the orthogonal transformation ${ O}^T$ to the un-primed axes; and in (\ref{boost}) the primed reference frame has its spatial axes  moving with velocity $-{\bf u}$ relative to the un-primed axes. 

The most general possible transformation between two IRFs is a combination of the 4 types shown above. This set of transformations forms a group -- the {\it Galilei group} -- with respect to composition of the transformations.  To specify a group transformation requires ten parameters $(a, {\bf b}, { O}, {\bf  u})$ (the set of orthogonal transformations $O$ are labeled by 3 parameters).  

The Galilean relativity principle is often stated as: ``the laws of physics do not depend upon the choice of inertial reference frame''.   One interpretation of this slogan is as follows.  A {\it closed} mechanical system is one that does not interact with its environment.  The existence of closed systems is, of course, an idealization, but it is a useful one.  The behavior of a closed system is not tied to any external elements which themselves may define a ``preferred'' reference frame ({\it e.g.} the rest frame of some laboratory equipment).  Therefore the behavior of a closed system should not be sensitive to the choice of inertial frame, that is to say, the mathematical description of the motion should in some sense look the same in all inertial reference frames.\footnote{If one wants to work with open systems which interact with some external environment, then under a change of reference frame one must not only transform the system but also one must take account of how the environment changes under the change of reference frame \cite{Brown1999}.  For simplicity in our subsequent exposition  we will focus on closed systems.}
Mathematically, given a solution to the equations of motion, applying any of the transformations (\ref{timetrans})--(\ref{boost}) to the solution should yield another solution to the {\it same} equations of motion.

\section{Galilei group in the Lagrangian formalism}
\label{Lagrange}
The Lagrangian for the non-relativistic Newtonian mechanics of $n$ particles is a function of positions and velocities, ${\bf r}_j$, ${\bf v}_j$, $j=1,2,\dots, n$, and time $t$,
\begin{equation}
L  = L({\bf r}_1, {\bf r}_2,\dots, {\bf r}_n, {\bf v}_1, {\bf v}_2,\dots, {\bf v}_n, t).
\label{genL}
\end{equation}
The equations of motion  of the system, which determine curves ${\bf r}_j = {\bf r}_j(t)$ in the configuration space, are   the Euler-Lagrange equations:
\begin{equation}
\left(\frac{\partial L}{\partial {\bf r}_j}\right)_{{\bf r} = {\bf r}(t)\ \ \atop {\bf v} = \frac{d}{dt}{\bf r}(t)} - \frac{d}{dt} \left(\frac{\partial L}{\partial {\bf v}_j}\right)_{{\bf r} = {\bf r}(t)\ \ \atop {\bf v} = \frac{d}{dt}{\bf r}(t)} = 0,\quad i = 1,2,\dots, n.
\label{ELeqns}
\end{equation}

The Lagrangian formulation of the laws of motion is equivalent to Newton's Galilean invariant form for a closed system with conservative forces if the positions of the particles are in a Euclidean space and the Lagrangian is chosen in the form
\begin{equation}
L({\bf r}_1,\dots, {\bf r}_n, {\bf v}_1,\dots, {\bf v}_n) = \sum_j \frac{1}{2} m_j v_j^2 - U({\bf r}_1,\dots, {\bf r}_n),
\label{NewtLag}
\end{equation}
where $v_j^2 = {\bf v}_j \cdot {\bf v}_j$.
Here the potential energy function is chosen to be Galilean invariant; it can only depend upon the relative locations of the particles using the Euclidean inner product.
The Galilei group of transformations is then a symmetry group of the equations of motion in the sense that the transformations (\ref{timetrans})--(\ref{boost}), when applied to a solution ${\bf r}_j(t), j=1,\dots, n$, of the equations of motion, will yield another solution, ${\bf r}_j^{\, \prime}(t), j=1,\dots, n$,
 of the same equations of motion.

It often happens that symmetries of equations of motion correspond to symmetries of the underlying Lagrangian in the sense that the Lagrangian is unchanged by the symmetry transformations.  Indeed, a sufficient condition for a transformation to be a symmetry of some Euler-Lagrange equations is that the transformation does not change the Lagrangian \cite{Arnold1989}.   The Lagrangian (\ref{NewtLag}) is invariant with respect to time translations because the kinetic and potential energies  are functions of ${\bf r}_j$ and ${\bf v}_j$ only. It is invariant with respect to spatial translations because the potential only depends upon scalars built from relative locations in a Euclidean invariant manner, and the kinetic energy only depends upon velocities (which are unchanged under spatial translations). The Lagrangian is invariant with respect to orthogonal transformations because it is constructed purely in terms of scalars built from the Euclidean scalar product. However, a boost will change the Lagrangian (\ref{NewtLag}) because of the kinetic energy term.  For the boosts we have
\begin{equation}
{\bf r}_j = {\bf  r}_j^{\, \prime} - t\,{\bf u},\quad {\bf v}_j = {\bf  v}_j^{\, \prime} - {\bf u},
\end{equation}
so we consider the effect on the Lagrangian  of transforming
\begin{equation}
{\bf  r}_j \to   {\bf r}_j - t\,{\bf u},\quad {\bf v}_j\to  {\bf v}_j - {\bf u}.
\label{boostJet}
\end{equation}
We get
\begin{align}
L(&{\bf r}_1,\dots, {\bf r}_n, {\bf v}_1,\dots, {\bf v}_n) \to  L({\bf r}_1-t\,{\bf u},\dots, {\bf r}_n-t\,{\bf u}, {\bf v}_1 - {\bf u},\dots, {\bf v}_n-{\bf u})\nonumber\\
&\equiv L^\prime({\bf r}_1,\dots, {\bf r}_n, {\bf v}_1,\dots {\bf v}_n)\nonumber\\
&= L({\bf r}_1,\dots, {\bf r}_n, {\bf v}_1,\dots, {\bf v}_n) + \sum_j m_j \left(\frac{1}{2} u^2-
{\bf u}\cdot {\bf v}_j \right) \nonumber\\
&= L({\bf r}_1,\dots, {\bf r}_n, {\bf v}_1,\dots, {\bf v}_n) + {\bf u}\cdot \left(\frac{1}{2} M {\bf u}-{\bf P} \right),
\label{boostL}
\end{align}
where $u^2 = {\bf u}\cdot {\bf u}$, $M$ is the total mass of the system, and ${\bf P}$ is the total mechanical momentum of the system: 
\begin{equation}
M = \sum_j m_j,\quad {\bf P} = \sum_j m_j {\bf v}_j .
\end{equation}
It might appear that one should use different Lagrangians in different IRFs, which would certainly not be in accord with the relativity principle! 
Fortunately, one {\it can} use the same Lagrangian (\ref{NewtLag}) in {\it any} IRF due to a fundamental feature of Lagrangian mechanics: two Lagrangians  will yield identical Euler-Lagrange equations if and only if they differ by the time derivative of a function.\footnote{Here we assume that the configuration space is Euclidean in its topology; otherwise there can be subtleties in this result \cite{Anderson1991}.}  Two Lagrangians which differ by a total time derivative are called {\it equivalent Lagrangians}. For the boosts, the transformed Lagrangian $L^\prime$ in eq.~(\ref{boostL}) differs from the original Lagrangian by a time derivative:
\begin{align}
L^\prime - L &=  {\bf u}\cdot \left(\frac{1}{2} M {\bf u}-{\bf P} \right)\nonumber\\ 
&= \frac{d}{dt} \left[{\bf u}\cdot\left(\frac{1}{2} t M {\bf u}-\sum_j m_j {\bf r}_j \right)\right]\nonumber\\
&=  \frac{d}{dt} \left[M{\bf u}\cdot\left(\frac{1}{2} t {\bf u}-{\bf R} \right)\right],
\label{divsymmboost}
 \end{align}
 where ${\bf R} = \sum_j m_j {\bf r}_j/M$ is the center of mass of the system and it is understood that ${d{\bf r}_j}/{dt} = {\bf v}_j$.  
Consequently, the Lagrangians $L$ and $L^\prime$ are equivalent; that is, they yield identical equations of motion.  One is thus free to use the Lagrangian (\ref{NewtLag}) in any reference frame. 

Transformations that preserve a given Lagrangian  are, naturally enough, called {\it Lagrangian symmetries}.  For example, the space-time translations and orthogonal transformations from the Galilei group are Lagrangian symmetries of eq.~(\ref{NewtLag}).  Transformations that only preserve a given Lagrangian up to addition of a time derivative of a function are  called {\it divergence symmetries}.\footnote{This terminology stems from the fact that in a general variational problem, where the number of independent variables can be greater than one, two Lagrangians are equivalent if they differ by the divergence of a vector field, which is the higher-dimensional generalization of the total time derivative of a function.}  Boosts are divergence symmetries of the Lagrangian (\ref{NewtLag}). From the point of view of classical mechanics, Lagrangian symmetries and divergence symmetries play essentially identical roles.  They both indicate the transformation in question is a symmetry of the equations of motion; that is, the transformation maps solutions to solutions, and they both indicate -- via Noether's theorem -- the existence of  conservation laws \cite{Olver1986}.  

Given a Lagrangian that is invariant under a symmetry transformation,  an equivalent Lagrangian will only admit the transformation as a divergence symmetry if the two Lagrangians differ by the time derivative of a function that is not invariant under the transformation.  Does the existence of divergence symmetries simply reflect a less than optimal choice of Lagrangian within the set of equivalent Lagrangians?  In the case of the boosts it {\it is} possible to find a Lagrangian equivalent to (\ref{NewtLag}) which is invariant under boosts. For example, consider the following  Lagrangian (apparently first displayed by Noether) \cite{ Azcarraga1995, Olver2022}:
\begin{equation}
\tilde L =  \sum_j \frac{1}{2} m_j \left({\bf v}_j - \frac{1}{t}{\bf r}_j\right)\cdot \left({\bf v}_j - \frac{1}{t}{\bf r}_j\right)  - U({\bf r}_1,{\bf r}_2, \dots, {\bf r}_n).
\label{Ltilde}
\end{equation}
It is easy to check that $\tilde L$ is unchanged by the boosts (\ref{boostJet}). $\tilde L$ is equivalent to $L$ because
\begin{equation}
\tilde L - L = \sum_j \frac{1}{2t} m_j\left(\frac{1}{t}r_j^2-2{\bf v}_j\cdot {\bf r}_j \right) = \frac{d}{dt} \left(- \sum_j \frac{1}{2t} m_j r_j^2\right),
\label{boostdiv}
\end{equation}
where $r_j^2 = {\bf r}_j \cdot {\bf r}_j$.   While $\tilde L$ is invariant under boosts, it is not invariant with respect to space-time translations -- these are now divergence symmetries.\footnote{An additional feature of this Lagrangian is that it is not defined at $t=0$, which is the conventional time chosen to signify when the origins of the two boost-related IRFs are coincident.  This time can be chosen to be arbitrarily far in the past if desired.}

It is natural to ask: is there a Lagrangian equivalent to (\ref{NewtLag}) that is invariant under the entire Galilei group?  It turns out  there is no such Lagrangian \cite{Azcarraga1995, Olver2022}.   To prove this one uses results from Lie algebra cohomology (see  Appendix \ref{Olver}). Any Lagrangian for Newtonian mechanics must admit either translations or boosts as divergence symmetries.  If one prefers manifest symmetry with respect to space-time translations and orthogonal transformations, then one can use (\ref{NewtLag}), in which the boost symmetry is not manifest.  Alternatively, using (\ref{Ltilde}), one can have manifest symmetry with respect to boosts and orthogonal transformations, but not with respect to space-time translations.  In what follows we will restrict attention to the Lagrangians (\ref{NewtLag}) and (\ref{Ltilde}), but the same sort of analysis would pertain to any equivalent Lagrangian.

\section{The Galilei group in Feynman's path integral formulation of quantum mechanics}

In this section we use the transformation of the Lagrangian (\ref{NewtLag}) with respect to boosts, in conjunction with Feynman's path integral construction of solutions to the \Schrodinger equation, to derive the non-trivial transformation of the wave function with respect to boosts \cite{Wigner1952, Bargmann1954, Hamermesh1960, Levy1963, Brown1999, Landau1977, Merzbacher1998,Ballentine2014}. In the next section we will see how to handle a Lagrangian such as (\ref{Ltilde}).


 The Hamiltonian corresponding to the Lagrangian (\ref{NewtLag}) is
\begin{equation}
H({\bf r}_1,\dots, {\bf r}_n, {\bf p}_1,\dots, {\bf p}_n) = \sum_j \frac{p_j^2}{2m_j} + U({\bf r}_1,\dots, {\bf r}_n),
\label{Ham0}
\end{equation}
where $p_j^2 = {\bf p}_j \cdot {\bf p}_j$.  As before, we shall limit our attention to closed systems that  have a Galilean-invariant potential energy function.  
The \Schrodinger equation associated to the Hamiltonian (\ref{Ham0}) in the position representation is
\begin{equation}
\hat H\psi \equiv -\sum_j \frac{\hbar^2}{2m_j}\nabla^2_j \psi + U({\bf r}_1, \dots, {\bf r}_n) \psi = i\hbar \frac{\partial\psi}{\partial t} .
\label{SE1}
\end{equation}
Here $\psi = \psi({\bf r}_1, \dots, {\bf r}_n, t)$ and $\nabla_j^2$ is the Laplacian in the coordinates of the $j^{th}$ particle.

Given an initial wave function $\psi_1({\bf r}_1,\dots,{\bf r}_n)$, the wave function at time $t$ satisfying  (\ref{SE1}) and matching $\psi_1$ when $t=t_1$ can be computed via
\begin{equation}
\psi({\bf r}_1,\dots,{\bf r}_n, t) = \left(\prod_{j=1}^n \int  d^3 r_j^{\,\prime}\right) 
 K({\bf r}_1,\dots,{\bf r}_n; {\bf r}_1^{\,\prime},\dots,{\bf r}_n^{\,\prime}; t, t_1)
 \psi_1({\bf r}_1^{\,\prime},\dots,{\bf r}_n^{\,\prime}),
 \label{psiKpsi1}
\end{equation}
where the integral is over the configuration space. In terms of the Hamiltonian operator $\hat H$ defined in (\ref{SE1}) and eigenvectors $|{\bf r}_1, \dots, {\bf r}_n\rangle$ of the position operators $\hat{\bf r}_j$,\footnote{The position eigenvectors are ``delta function normalized'':
\begin{equation}
\langle {\bf r}_1^\prime, \dots, {\bf r}_n^\prime|{\bf r}_1, \dots, {\bf r}_n\rangle = \prod_{j=1}^n \delta^3({\bf r}_j^\prime - {\bf r}_j),\  {\rm where}\  \delta^3({\bf r}_j^\prime - {\bf r}_j) = \delta(x_j^\prime - x_j)\delta(y_j^\prime - y_j) \delta(z_j^\prime - z_j).\nonumber
\end{equation}
}
 we have
\begin{equation}
K({\bf r}_1,\dots,{\bf r}_n, {\bf r}_1^{\,\prime},\dots,{\bf r}_n^{\,\prime}; t, t_1) 
=\langle {\bf r}_1,\dots,{\bf r}_n| e^{-\frac{i}{\hbar}(t-t_1) \hat H}|
{\bf r}_1^{\,\prime},\dots,{\bf r}_n^{\,\prime}\rangle.
\label{TevolK}
\end{equation} 
Here we have introduced the fundamental solution $K$ of the Schr\"odinger equation,   also called the ``propagation kernel'', or just the ``propagator''. For a time-independent Hamiltonian it is given by matrix elements of the exponential of the Hamiltonian \cite{Merzbacher1998, Sakurai2020}, as shown in (\ref{TevolK}).  The formula (\ref{psiKpsi1}) yields the solution to the \Schrodinger equation  (\ref{SE1}) with initial wave function $\psi_1$ by virtue of the fact that the propagator $K$ satisfies
\begin{align}
\left(-\sum_j \frac{\hbar^2}{2m_j}\nabla^2_j  + U({\bf r}_1, \dots, {\bf r}_n)\right) &K({\bf r}_1,\dots,{\bf r}_n, {\bf r}_1^{\,\prime},\dots,{\bf r}_n^{\,\prime}; t, t_1)\nonumber\\  
&= i\hbar \frac{\partial}{\partial t} K({\bf r}_1,\dots,{\bf r}_n, {\bf r}_1^{\,\prime},\dots,{\bf r}_n^{\,\prime}; t, t_1) ,
\label{SchrProp1}
\end{align}
and
\begin{equation}
K({\bf r}_1,\dots,{\bf r}_n, {\bf r}_1^{\,\prime},\dots,{\bf r}_n^{\,\prime}; t_1, t_1)  = \prod_{j=1}^n \delta^3({\bf r}_j - {\bf r}_j^{\, \prime}).
\label{SchrProp2}
\end{equation}

The propagator can be defined and/or computed via Feynman's path integral:
\begin{equation}
K({\bf r}_1,\dots,{\bf r}_n, {\bf r}_1^{\,\prime},\dots,{\bf r}_n^{\,\prime}; t_2, t_1)   = \int_{{\bf r}^{\, \prime},\, t_1}^{{\bf r}^{\, \prime\prime}, \, t_2} \D {\bf r}\, e^{\frac{i}{\hbar} S[{\bf r}]}.
\label{FPI}
\end{equation}
The details of the somewhat intricate definition of this ``integral'' in terms of a limit of integrals over discretized paths are available in many articles and texts; see, {\it e.g.}, references \cite{Feynman1965, Schulman1981, Marinov1980, Merzbacher1998, Sakurai2020}.  For our purposes, the main features are as follows. The paths are in the configuration space of the system and they begin at ${\bf r}_1^{\,\prime},\dots, {\bf r}_n^{\,\prime}$ when $t=t_1$ and they end at ${\bf r}_1^{\,\prime\prime},\dots,{\bf r}_n^{\,\prime\prime}$ when  $t=t_2$.  For each path  in configuration space, $\left({\bf r}_1(t), \dots, {\bf r}_n(t)\right)$, the quantity $S$ is the action integral
\begin{equation}
S[{\bf r}_1,\dots,{\bf r}_n] = \int_{t_1}^{t_{2}}dt\,  L,
\end{equation}
evaluated on the path. The ``integration'' in (\ref{FPI}) is defined by partitioning the time interval into $N+1$ steps of size $\epsilon =\frac{t_2-t_1}{N+1}$ and discretizing the paths of each particle into continuous, piecewise linear paths connecting the points ${\bf r}_{(k)} = {\bf r}(t_{(k)})$, $(k) = (0), (1),\dots, (N+1)$, where $ {\bf r}_{(0)} = {\bf r}^\prime$ and $ {\bf r}_{(N+1)} = {\bf r}^{\prime\prime}$.\footnote{In order to keep the notation palatable, here we suppress the label denoting the particles, {\it e.g.,} subscript $j=1,2,\dots, n$, in favor of the label for the time steps, {\it e.g.,} subscript $(k)= (0), (1), \dots, (N+1)$.}   The action integral is defined by a Riemann sum $S_N$ using the discretization. Finally, the integration measure for finite $N$ is given by the product over all particles of
\begin{equation}
\int \left[{\cal D}{\bf r}\right]_N \equiv  \int d^3 r_{(1)}\cdots \int d^3 r_{(N)} \left(\frac{m}{2\pi i \hbar\epsilon}\right)^{3(N+1)/2},
\label{dismeas}
\end{equation}
with the path integral itself defined by
\begin{equation}
\int_{{\bf r}^{\, \prime},\, t_1}^{{\bf r}^{\, \prime\prime}, \, t_2} \D {\bf r}\, e^{\frac{i}{\hbar} S[{\bf r}]} = \lim_{N\to\infty} \left[\left(\prod_{\rm particles}\int \left[{\cal D}{\bf r}\right]_N\right) e^{\frac{i}{\hbar} S_N}\right].
\end{equation}



We will now derive an identity which characterizes the transformation of the propagator under boosts. This identity will imply a corresponding transformation law for the wave function. Consider
\begin{equation}
K({\bf r}^{\, \prime\prime} - t_2{\bf u} , {\bf r}^{\, \prime} - t_1 {\bf u} , t_2, t_1) = \int_{{\bf r}^{\, \prime} - t_1{\bf u},\,  t_1}^{{\bf r}^{\, \prime\prime} -t_2 {\bf u}, \, t_2} \D {\bf r}\,e^{\frac{i}{\hbar} S[{\bf r}]}.
\end{equation}
Note that the paths begin at ${\bf r}(t_1) = {\bf r}^{\, \prime} - t_1{\bf u}$ and they end at ${\bf r}(t_2) = {\bf r}^{\, \prime\prime} - t_2{\bf u}$. Change integration variables:
\begin{equation}
{\bf r}_j(t) = {\bf y}_j(t) - t\,{\bf u},\quad j = 1,2,\dots, n.
\end{equation}
The discrete version of this change of variables is (for each particle) a translation:
\begin{equation}
{\bf r}_{(k)} = {\bf y}_{(k)} - t_{(k)} {\bf u}.
\end{equation}
The discretized integration measures in eq.~(\ref{dismeas}) are translationally invariant: $d^3 r_{(k)} = d^3 y_{(k)}$ so that, formally, ${\cal D}{\bf r}\, = {\cal D}{\bf y}\,$.  In terms of the new variables the end-points of the paths are 
\begin{equation}
{\bf y}(t_1) = {\bf r}(t_1) + t_1 {\bf u} = {\bf r}^{\, \prime},\quad {\bf y}(t_2) = {\bf r}(t_2) + t_2{\bf u} = {\bf r}^{\, \prime\prime}.  
\end{equation}
In terms of the new integration variables the action is (see (\ref{divsymmboost}))
\begin{equation}
S[{\bf r}] = S[{\bf y}] + \beta(t_2) - \beta(t_1),
\end{equation}
where
\begin{equation}
\beta(t) = M{\bf u}\cdot\left(\frac{t}{2}  {\bf u}-{\bf Y}(t) \right)
\end{equation}
and ${\bf Y} = \sum_j m_j {\bf y}_j /M$.

The path integral therefore takes the form
\begin{align}
K({\bf r}^{\, \prime\prime} - t_2{\bf u} , {\bf r}^{\, \prime} - t_1 {\bf u} , t_2, t_1) &= \int_{{\bf r}^{\, \prime},\,  t_1}^{{\bf r}^{\, \prime\prime},\,  t_2} \D {\bf y}\, e^{\frac{i}{\hbar} S[{\bf y} - t{\bf u}]}
\nonumber\\
&=
e^{\frac{i}{\hbar}\left\{\beta(t_2) - \beta(t_1)\right\} }
\int_{{\bf r}^{\, \prime}, \, t_1}^{{\bf r}^{\, \prime\prime}, \, t_2} \D {\bf  y}\, e^{\frac{i}{\hbar} S[{\bf y}] }\nonumber\\
&=e^{\frac{i}{\hbar}\left\{\beta(t_2) - \beta(t_1)\right\} }
 K({\bf r}^{\, \prime\prime}, {\bf r}^{\, \prime}, t_2, t_1).
\end{align}
Thus we get the identity
\begin{equation}
K({\bf r}^{\, \prime\prime} - t_2{\bf u}, {\bf r}^{\, \prime} - t_1{\bf u} , t_2, t_1) = e^{\frac{i}{\hbar} M{\bf u}\cdot (t_2{\bf u}/2 - {\bf R}^{\prime\prime})} K({\bf r}^{\, \prime\prime}, {\bf r}^{\, \prime}, t_2, t_1)e^{-\frac{i}{\hbar} M{\bf u}\cdot (t_1{\bf u}/2 - {\bf R}^{\prime})}.
\label{Kid}
\end{equation}
We used the transformation properties of the Lagrangian with respect to boosts to derive this identity.  It can, of course, also be obtained from the usual operator formalism of quantum mechanics via eq.~(\ref{TevolK}).
 
 The identity
 (\ref{Kid}) implies the transformation rule for the wave function as follows.  
 Let the wave function at time $t_1$ be $\psi_1({\bf r}_1,\dots, {\bf r}_n) \equiv \psi_1({\bf r})$.  At a later time $t$ the wave function satisfying (\ref{SE1}) is 
 \begin{equation}
\psi({\bf r}_1, \dots, {\bf r}_n, t) \equiv  \psi({\bf r}, t) = \left(\prod_{j=1}^n \int d^3r_j^\prime\right) K({\bf r}, {\bf r}^{\, \prime}, t, t_1) \psi_1({\bf r}^{\, \prime}).
 \end{equation}
 Let ${\bf r}_j = {\bf y}_j - t\,{\bf u}$, ${\bf r}_j^{\, \prime} ={\bf y}_j^{\, \prime}- t_1{\bf u}$ for all $j=1,2,\dots,n$; using (\ref{Kid}) we have
 \begin{align}
 \psi({\bf y} - t\,{\bf u}, t) &= \left(\prod_{j=1}^n \int d^3y_j^\prime\right) K({\bf y} - t{\bf u}, {\bf y}^{\,\prime} - t_1{\bf u}, t, t_1) \psi_1({\bf y}^{\,\prime} - t_1{\bf u})\nonumber\\
 &=e^{\frac{i}{\hbar} \beta(t)}  \left(\prod_{j=1}^n \int d^3y_j^\prime\right)K({\bf y}, {\bf y}^{\, \prime}, t, t_1)e^{-\frac{i}{\hbar} \beta(t_1)}\psi_1({\bf y}^{\,\prime} - t_1{\bf u})
 \label{boostident0}
 \end{align}
 or
\begin{equation}
e^{-\frac{i}{\hbar} \beta(t)} \psi({\bf y} - t\,{\bf u}, t) 
= \left(\prod_{j=1}^n \int d^3y_j^\prime\right)K({\bf y}, {\bf y}^{\, \prime}, t, t_1)e^{-\frac{i}{\hbar} \beta(t_1)}\psi_1({\bf y}^{\,\prime} - t_1{\bf u}).
\label{boostident}
\end{equation}
We can write (\ref{boostident}) as
\begin{equation}
\psi_{\bf u}({\bf r}, t) = \left(\prod_{j=1}^n \int d^3r_j^\prime\right)K({\bf r}, {\bf r}^{\, \prime}, t, t_1) \psi_{{\bf u}}({\bf r}^\prime, t_1),
\end{equation}
where
\begin{equation}
\psi_{{\bf u}}({\bf r}, t)  = e^{-\frac{i}{\hbar} M{\bf u}\cdot (t\,{\bf u}/2 - {\bf R})}  \psi({\bf r} - t\,{\bf u}, t).
\label{psitransboost}
\end{equation}
Because the propagator  satisfies the \Schrodinger equation (see (\ref{SchrProp1})), it follows that $\psi_{{\bf u}}({\bf r}, t)$ solves the same \Schrodinger equation  for any choice of ${\bf u}$.  Because the propagator satisfies the initial conditions (\ref{SchrProp2}), the initial wave function for $\psi_{\bf u}$ is 
\begin{equation}
\psi_{1{\bf u}} = e^{-\frac{i}{\hbar} M{\bf u}\cdot (t_1\,{\bf u}/2 - {\bf R})}  \psi_1({\bf r} - t_1\,{\bf u}).
\end{equation}
Put differently, if $\psi({\bf r}, t)$ satisfies the \Schrodinger equation (\ref{SE1}) with initial wave function $\psi_1({\bf r})$, then   $\psi_{\bf u}({\bf r}, t)$ solves the same \Schrodinger equation with initial wave function $\psi_{1\bf u}({\bf r})$.


The transformation law (\ref{psitransboost}) is normally derived in texts (see, e.g., \cite{Merzbacher1998, Ballentine2014}) by changing variables ${\bf r}_j \to {\bf r}_j -t\,{\bf u}$ in the Schr\"odinger equation and looking for a transformation law for the wave function which yields a solution in the new IRF.  A more systematic method for deriving eq.~(\ref{psitransboost}) is to compute the point symmetries of the Schr\"odinger equation \cite{Olver1986}.

Viewing the transformation (\ref{psitransboost}) from the ``active'' point of view, if $\psi$ is the state of the system, then $\psi_{{\bf u}}$ is the state that would be obtained if the {\it system} were boosted with  velocity ${\bf u}$.  For example, if the absolute value of the wave function is peaked about some points ${\bf r}_j = {\bf r}_{0j}$, then the transformed wave function will be peaked about the points ${\bf r}_j = {\bf r}_{0j} + t\,{\bf u}$.
From the ``passive'' point of view of two IRFs, if a system is described by a wave function $\psi$ in IRF-1, then $\psi_{{\bf u}}$ is the form of the wave function in a transformed frame, IRF-2, which is moving with  velocity $-{\bf u}$ relative to  IRF-1. 

The argument ${\bf r}_j - t {\bf u}$, $j = 1,2,\dots, n$, of the wave function in eq.~(\ref{psitransboost}) is to be expected, but the necessity for the multiplicative  factor is not obvious and, as pointed out in references \cite{Brown1999, Ballentine2014}, emphasizes the fact that the wave function cannot be viewed in terms of a physical wave propagating in space because the latter would transform without the multiplicative factor.  
The transformation rule (\ref{psitransboost}) is analogous to the transformation law of a vector field ${\bf V}({\bf r})$ under an orthogonal transformation ${ O}$ of the underlying Euclidean space:
\begin{equation}
{\bf V}({\bf r}\,) \to { O}\cdot {\bf V}({ O}^{-1}\cdot {\bf r}\,).
\end{equation}
The directions of the vectors making up the vector field are transformed along with the locations of the vectors.  Compare this with the transformation rule of a scalar field, {\it i.e.,} a function, where only the location of the scalar is transformed.
  
Finally, consider  the Galilei subgroup of space-time translations and rotations. These are Lagrangian symmetries so the action integral is not changed by these transformations. The discretized integration measures (\ref{dismeas}), consisting of products of the ordinary volume elements for Euclidean space, are easily checked to be  invariant under translations and rotations. Using the same analysis as above,  it  follows that for this subgroup the wave function transforms ``as a scalar'', that is, by simply transforming its argument.  In particular, if $\psi({\bf r}_1,\dots, {\bf r}_n, t)$ solves the Schr\"odinger equation (\ref{SE1}) then so does
\begin{equation}
\psi({\bf r}_1,\dots, {\bf r}_n, t-a),\quad   \psi({\bf r}_1 - {\bf b}, \dots, {\bf r}_n - {\bf b}, t),\quad
\psi({ O}^T\cdot {\bf r}_1,\dots, { O}^T\cdot {\bf r}_n,t)
\end{equation}
with corresponding interpretations from the active and passive points of view.

\section{The path integral using the boost-invariant Lagrangian (\ref{Ltilde})}
 
The same sort of analysis used in the previous section can be used when the Lagrangian is (\ref{Ltilde}).  The path integral is defined as before.\footnote{However, as explained in \cite{Schulman1981}, the terms in the Lagrangian proportional to ${\bf r}\cdot {\bf v}$ should be discretized using the midpoint rule:
\begin{equation}
{\bf r}\cdot {\bf v} \quad\longrightarrow\quad \frac{1}{2}({\bf r}_{(k)} + {\bf r}_{(k+1)})\cdot({\bf r}_{(k+1)} - {\bf r}_{(k)} )\nonumber
\end{equation}
}
The discretized integration measure (\ref{dismeas}) consists of products of   volume elements for Euclidean space and so is Galilean invariant.
The Lagrangian (\ref{Ltilde}) is invariant with respect to boosts and rotations so the wave function will transform as a scalar field with respect to this subgroup of transformations. Because space and time translations are now divergence symmetries, a non-trivial transformation law for the wave function will arise for these transformations.  

Let us  construct the \Schrodinger equation that the path integral is designed to solve. The  Hamiltonian corresponding to the Lagrangian (\ref{Ltilde}) is 
\begin{equation}
\tilde H = \sum_j \left(\frac{p_j^2}{2m_j} + \frac{1}{t}{\bf r}_j\cdot {\bf p}_j\right) + U({\bf r}_1,\dots, {\bf r}_n).
\label{Htilde}
\end{equation}
This result can be obtained as follows.  The momentum ${\bf p}_j$ conjugate to ${\bf r}_j$ is given by
\begin{equation}
 {\bf  p}_j = m_j\left({\bf v}_j - \frac{1}{t}{\bf r}_j\right)\quad \longleftrightarrow\quad {\bf v_j} = \frac{1}{m}_j {\bf   p}_j + \frac{1}{t} {\bf r}_j.
\end{equation}
The Legendre transformation defining the Hamiltonian $\tilde H$ is
\begin{align}
\tilde H &= \sum_j {\bf   p}_j \cdot {\bf v}_j - \tilde L\nonumber\\
&= \sum_j \left\{{\bf   p}_j \cdot \left(\frac{1}{m}_j {\bf   p}_j + \frac{1}{t} {\bf r}_j\right) - \frac{1}{2 m_j} {\bf   p}_j\cdot {\bf   p}_j\right\} + U({\bf r}_1, {\bf r}_2,\dots, {\bf r}_n)\nonumber\\
&= \sum_j \left(\frac{p_j^2}{2m_j} + \frac{1}{t}{\bf r}_j\cdot {\bf p}_j\right) + U({\bf r}_1,\dots, {\bf r}_n).
\end{align}
Alternatively, the Hamiltonian $\tilde H$ in (\ref{Htilde}) can  be obtained starting from (\ref{Ham0}) by making the {\it time dependent} canonical transformation $({\bf r}_j, {\bf p}_j) \to  ({\bf r}_j,  {\bf p}_j - (m_j/t ){\bf r}_j)$, provided one takes account of the change in the Hamiltonian required by the time dependence of the transformation \cite{Landau1976}.

The  Schr\"odinger equation corresponding to (\ref{Htilde}) is
\begin{equation}
\sum_j\left\{ -\frac{\hbar^2}{2m_j}\nabla_{j}^2 \psi + \frac{\hbar}{2it} \left[{\bf r}_j\cdot \nabla_j\psi + \nabla_{j}\cdot ({\bf r}_j \psi) \right]\right\}+ U\psi = i\hbar \frac{\partial\psi}{\partial t}.
\label{tildeSchrod}
\end{equation}
We have symmetrized the product of position and momentum operators to ensure the Hamiltonian is Hermitian.
The correspondence between the Schr\"odinger equations (\ref{tildeSchrod}) and (\ref{SE1}) is as follows. If $\psi$ solves (\ref{SE1}) then 
\begin{equation}
 \tilde \psi = \exp\left\{-\frac{i}{\hbar} \sum_j \frac{1}{2t} m_j r_j^2\right\}\psi
\end{equation}
 solves (\ref{tildeSchrod}).  This result is straightforward to check directly. It can also be understood from the path integral expression (\ref{FPI}) for solutions to the \Schrodinger equation by using the relationship between $L$ and $\tilde L$ given in eq.~(\ref{boostdiv}).
  
Following the analysis of the previous section, we now use the transformation properties of the Lagrangian with respect to space and time translations to deduce the transformation law for the wave function with respect to these changes in the inertial reference frame. The change in the Lagrangian (\ref{Ltilde}) under space translations, ${\bf r}_j \to {\bf r}_j - {\bf b}$, is 
\begin{equation}
\tilde L({\bf r}_1 - {\bf b},\dots, {\bf r}_n - {\bf b}, {\bf v}_1,\dots, {\bf v}_n, t) = \tilde L({\bf r}_1,\dots, {\bf r}_n, {\bf v}_1, \dots, {\bf v_n}) + \frac{d}{dt}\left\{\frac{M}{t}{\bf b}\cdot \left({\bf R}- \frac{1}{2}{\bf b}\right)\right\}.
\end{equation}
This implies the propagator identity:
\begin{equation}
K({\bf r}^{\, \prime\prime} - {\bf b}, {\bf r}^{\, \prime} - {\bf b}, t_2, t_1) = e^{\frac{i}{\hbar}\sigma(t_2)}
K({\bf r}^{\, \prime\prime}, {\bf r}^{\, \prime}, t_2, t_1)e^{-\frac{i}{\hbar}
\sigma(t_1)},
\end{equation}
where
\begin{equation}
\sigma(t) = \frac{M}{t}{\bf b}\cdot \left({\bf R}(t)- \frac{1}{2}{\bf b}\right).
\end{equation}
The corresponding transformation of the wave function is
\begin{equation}
\psi_{{\bf b}}({\bf r}_1,\dots, {\bf r}_n, t) = \exp\left\{-\frac{i}{\hbar}\frac{M}{t}{\bf b}\cdot \left({\bf R}- \frac{1}{2}{\bf b}\right)\right\}\psi({\bf r}_1- {\bf b},\dots, {\bf r}_n - {\bf b}, t).
\label{Spsi}
\end{equation}
It is straightforward (if somewhat tedious) to check that if $\psi({\bf r}_1,\dots, {\bf r}_n, t)$ solves the \Schrodinger equation (\ref{tildeSchrod}) then so does 
$\psi_{{\bf b}}({\bf r}_1,\dots, {\bf r}_n, t)$.

The change in the Lagrangian (\ref{Ltilde}) under time translations, $t \to t- a$, is 
\begin{equation}
\tilde L({\bf r}_1,\dots, {\bf r}_n, {\bf v}_1,\dots, {\bf v}_n, t - a) = 
\tilde L({\bf r}_1,\dots, {\bf r}_n, {\bf v}_1,\dots, {\bf v}_n, t) + \frac{d}{dt} \left\{-\frac{a}{2t(t-a)}\sum_j m_j r_j^2\right\}.
\end{equation}
This implies the propagator identity:
\begin{equation}
K({\bf r}^{\, \prime\prime}, {\bf r}^{\, \prime}, t_2-a, t_1-a) = e^{\frac{i}{\hbar}\tau(t_2)}K({\bf r}^{\, \prime\prime}, {\bf r}^{\, \prime}, t_2, t_1)e^{-\frac{i}{\hbar}\tau(t_1)},
\end{equation}
where
\begin{equation}
\tau(t) = -\frac{a}{2t(t-a)}\sum_j m_j r_j^2(t).
\end{equation}
and the corresponding transformation of the wave function is
\begin{equation}
\psi_{a}({\bf r}_1,\dots, {\bf r}_n, t) = \exp\left\{\frac{i}{\hbar}\frac{a}{2t(t-a)}\sum_j m_j r_j^2\right\}\psi({\bf r}_1,\dots, {\bf r}_n, t - a).
\label{Tpsi}
\end{equation}
It is straightforward to check that if $\psi({\bf r}_1,\dots, {\bf r}_n, t)$ solves the \Schrodinger equation (\ref{tildeSchrod}) then so does 
$\psi_{a}({\bf r}_1,\dots, {\bf r}_n, t)$.

\section{The projective representation of the Galilei group in quantum mechanics}
\label{QM2}

Given a group $G$ of transformations, a  {\it representation} of $G$ on a vector space ${\cal V}$ is an identification of a linear operator $L(g)$ on ${\cal V}$ for each element $g\in G$ such that 
 \begin{equation}
 L(g_1) L(g_2) = L(g_1g_2),
 \label{Rep}
\end{equation}
where the composition of linear operators is used on the left side of the equality and the group product on the right.  In quantum mechanics, the vector space is the  Hilbert space of wave functions and the linear operators are usually chosen to be unitary.  In the case of the Galilei group, the representation of the group on wave functions is intended to transform the state of a given physical system from the point of view of one IRF to that of  another  IRF.   

As an example of eq.~(\ref{Rep}) consider the Galilean subgroup of boosts with the corresponding transformation of wave functions as given in eq.~(\ref{psitransboost}):
\begin{equation}
L({\bf u}) \psi({\bf r}_1,\dots, {\bf r}_n, t) = \exp\left\{-\frac{i}{\hbar} M{\bf u}\cdot (\frac{t}{2}{\bf u} - {\bf R})\right\}  \psi({\bf r}_1 - t{\bf u}, \dots, {\bf r}_n - t{\bf u}, t) \equiv \psi_{{\bf u}}({\bf r}, t).
\label{Lboost}
\end{equation}
The composition of two boosts determined by ${\bf u}_1$ and ${\bf u}_2$ is another boost determined by ${\bf u}_1 + {\bf u}_2$:
\begin{equation}
{\bf r} \to {\bf r} + t{\bf u}_1 \to ({\bf r} + t{\bf u}_1) + t{\bf u}_2 = {\bf r} + t({\bf u}_1 + {\bf u}_2).
\end{equation}
From the definition (\ref{Lboost})  it follows that
\begin{equation}
 L({\bf u}_2)\left(L({\bf u}_1)  \psi\right) = \psi_{{\bf u}_1 + {\bf u}_2} =  L({\bf u}_1 + {\bf u}_2)\psi .
\end{equation}
Thus the wave functions transforming according to eq.~(\ref{Lboost}) satisfy the representation property for the subgroup of boosts. 

On the other hand, the behavior of the wave functions with respect to the subgroup of boosts and spatial translations only satisfies the representation property {\it up to a phase factor}.  To see what is meant here, consider the composition  of a boost determined by ${\bf u}$ followed by a spatial translation determined by ${\bf b}$:
\begin{equation}
{\bf r}  \to  {\bf r} + t{\bf u}  \to  {\bf r} + t{\bf u} + {\bf b}.
\end{equation}
If, instead, we first translate and then boost we get the same result:
\begin{equation}
{\bf r}  \to  {\bf r} +{\bf b}  \to  {\bf r} + {\bf b} +   t{\bf u}  = {\bf r} + t{\bf u} + {\bf b}.
\end{equation}
Evidently the boosts and translations commute.  The representation of spatial translations on wave functions from \S4 is
\begin{equation}
S({\bf b}) \psi({\bf r}_1,\dots, {\bf r}_n, t) = \psi({\bf r}_1 - {\bf b},\dots, {\bf r}_n - {\bf b}, t).
\end{equation}
Now consider the transformation of a wave function by a boost
$L({\bf u})$ with relative velocity ${\bf u}$ followed by a spatial translation $ S({\bf b})$ by the vector ${\bf b}$:
\begin{align}
 S({\bf b}) & L({\bf u}) \psi({\bf r}_1,\dots,{\bf r}_n, t) =  S({\bf b}) \psi_{{\bf u}}\nonumber\\ 
&= \exp\left\{\frac{i}{\hbar}\left[ \sum_j m_j({\bf r}_j - {\bf b})\cdot u - \frac{t}{2}Mu^2 \right]\right\}\psi({\bf r}_1 - t{\bf u} - {\bf b},\dots,{\bf r}_n - t{\bf u} -{\bf b}, t)\nonumber\\ 
&= \exp\left\{-\frac{i}{\hbar} M {\bf b}\cdot {\bf u}\right\}  L({\bf u}) S({\bf b})\cdot \psi({\bf r}_1,\dots,{\bf r}_n, t).
\label{SLphase}
\end{align}
The two operations fail to commute because of the appearance of the phase factor $e^{-iM{\bf b}\cdot {\bf u}/\hbar}$.  One says that in quantum mechanics the Galilei group is ``projectively represented''  or has a ``representation up to a phase''  \cite{Wigner1952, Bargmann1954, Hamermesh1960, Levy1963, Ballentine2014}.   
The appearance of the phase factor does not interfere with the physical interpretation of quantum mechanics with respect to changes of IRF because  phase factors are guaranteed to drop out when constructing any observable quantity.  Any two  wave functions that differ by a phase are  physically equivalent. For example, probability distributions will not be affected by such a phase factor.  Thus quantum mechanics can and does take advantage of this weakened notion of representation and one often views such projective representations as exhibiting intrinsically quantum mechanical behavior.\footnote{In this regard, a famous example of a projective representation occurs when studying the behavior of states of a spin 1/2 system ({\it e.g.}, an electron) under the group of rotations (see, {\it e.g.} references \cite{Merzbacher1998, Ballentine2014}).  It would be interesting to examine this example from the perspective of our exposition here.}  Indeed, according to \cite{Wigner1952, Bargmann1954, Hamermesh1960, Levy1963, Ballentine2014} the projective representation of the Galilei group is mandatory if one wants the quantum mechanical system to have physically acceptable properties. 

As an exercise, the interested reader can  repeat the same calculation using the framework of \S 5 where the wave function transforms as a scalar with respect to boosts and has  non-trivial transformations with respect to space-time translations.  See equations (\ref{Spsi}) and (\ref{Tpsi}).  The result is the same as   in eq.~(\ref{SLphase}):  the boosts and translations fail to commute and the subgroup of boosts and spatial translations is only represented up to a phase $e^{-iM{\bf b}\cdot {\bf u}/\hbar}$.    As we have seen, this immutable aspect of the projective representation  is neatly understood in terms of the transformation properties of the Lagrangian by virtue of Feynman's path integral formalism.  From a more mathematical perspective, the connection between the group representation  on wave functions and the transformation properties of the Lagrangian can be understood using the algebraic properties of the group of Galilean transformations.    One can read more about this in Appendices A and B.

\appendix

\section{Infinitesimal Galilean transformations and invariant Lagrangians}
\label{Olver}

Given a Lagrangian admitting a group of divergence symmetries, Olver has given necessary and sufficient conditions (in a very general setting) for the existence of an equivalent Lagrangian which admits the same group of transformations as Lagrangian symmetries \cite{Olver2022}.  Here we show how this works in the case of Galilean symmetry in Newtonian mechanics (see also ref. \cite{Azcarraga1995}).  We shall see that a relatively simple property of the Lie algebra of infinitesimal Galilean transformations guarantees that there is no Lagrangian for Newtonian mechanics which is strictly invariant under the Galilei group.

To understand Olver's condition in the context of our discussion we will use the infinitesimal form of the Galilei group of transformations.  The properties of the infinitesimal transformations we shall need are as follows.

If $(t, {\bf r})\to ( t^\prime, {{\bf r}}^{\, \prime})$ is a  family of transformations labeled by one of the parameters of the Galilei group then, for an infinitesimal value $\epsilon$ for the parameter, the infinitesimal change $(\delta t, \delta {\bf r})$ of the space-time variables is defined as
\begin{equation}
t^\prime = t + \epsilon\, \delta t + {\cal O}(\epsilon^2),\quad {{\bf r}}^{\, \prime} = {\bf r} + \epsilon\, \delta {\bf r} + {\cal O}(\epsilon^2).
\end{equation}
The time translations, space translations, rotations, and boosts defined in (\ref{timetrans})--(\ref{boost}) correspond to  the following infinitesimal changes in ${\bf r}$ and $t$:
\begin{equation}
\delta_0 t = 1,\quad \delta_0 {\bf r}  = 0,\quad {\rm time\ translations}, 
\label{delta0}
\end{equation}
\begin{equation}
\delta_1 t = 0,\quad \delta_1 {\bf r}  = {\bf i}, \quad \delta_2 t = 0,\quad \delta_2 {\bf r}  = {\bf j}, \quad \delta_3 t = 0,\quad \delta_3 {\bf r}  = {\bf k}, \quad {\rm space\ translations}, 
\label{delta1}
\end{equation}
\begin{equation}
\delta_4 t = 0,\quad \delta_4 {\bf r}  = R_1\cdot {\bf r} ,\quad \delta_5 t = 0,\quad \delta_5{\bf r}  = R_2\cdot {\bf r} , \quad\delta_6 t = 0,\quad  \delta_6 {\bf r}  = R_3\cdot {\bf r} ,\quad{\rm rotations,}
\label{delta4}
\end{equation}
\begin{equation}
\delta_7 t = 0,\quad \delta_7{\bf r}  =  t{\bf i}, \quad \delta_8 t = 0,\quad \delta_8{\bf r}  = t{\bf j}, \quad \delta_9 t = 0,\quad \delta_9{\bf r}  =  t{\bf k},\quad {\rm boosts.}
\label{delta9}
\end{equation}
Here ${\bf i}$, ${\bf j}$, ${\bf k}$ are the usual Cartesian coordinate unit basis vectors, and
\begin{equation}
R_1 = 
\begin{pmatrix}
0 & 0 & 0 \\
0 & 0 & 1\\
0 & -1 & 0\\
\end{pmatrix},
\quad
R_2 = 
\begin{pmatrix}
0 & 0 & -1 \\
0 & 0 & 0\\
1 & 0 & 0\\
\end{pmatrix},\quad
R_3 = 
\begin{pmatrix}
0 & 1 & 0 \\
-1 & 0 & 0\\
0 & 0 & 0\\
\end{pmatrix}.
\end{equation}
Any infinitesimal transformation $(\delta t, \delta {\bf r}\,)$ from the Galilei group can be expressed as a linear combination of the foregoing transformations.

The above equations (\ref{delta0})--(\ref{delta9})  define infinitesimal transformations  of any function $f = f({\bf r}, t)$ of space and time by transforming the argument of $f$:
\begin{equation}
f({\bf r} + \epsilon \delta {\bf r} + {\cal O}(\epsilon^2),  t + \epsilon \delta t + {\cal O}(\epsilon^2)) = f({\bf r}, t) + \epsilon\, \delta f({\bf r}, t) + {\cal O}(\epsilon^2).
\end{equation}
 These transformations satisfy the commutation relations
\begin{equation}
[\delta_\alpha, \delta_\beta] f({\bf r}, t) = \sum_{\gamma=0}^9C_{\alpha\beta}^\gamma \delta_\gamma f({\bf r}, t),\quad \alpha, \beta, = 0,1, 2,\dots, 9, 
\label{GalileiStructureEq}
\end{equation}
where 
\begin{align}
&C_{07}^1 = C_{08}^2 = C_{09}^3 = C_{15}^3 = - C_{16}^2 = -C_{24}^3 = C_{26}^1 = C_{34}^2 = -C_{35}^1=C_{45}^6\nonumber\\
& = - C_{46}^5 = C_{48}^9 = - C_{49}^8 = C_{56}^4 = - C_{57}^9 = C_{59}^7 = C_{67}^8 = -C_{68}^7 = 1,
\label{structure}
\end{align}
$C^\alpha_{\beta\gamma}=-C_{\gamma\beta}^\alpha$, and all other constants vanish. Thus the infinitesimal Galilean transformations (\ref{delta0})--(\ref{delta9}) form a Lie algebra \cite{Schutz1980} with structure constants $C^\alpha_{\beta\gamma}$.

The infinitesimal transformations of particle positions are given by applying eqs.~(\ref{delta0})--(\ref{delta9}) simultaneously to each ${\bf r}_j$, $j=1,2,\dots,n$.  The infinitesimal transformations of particle velocities are defined by
\begin{equation}
\delta_\alpha {\bf v}_j = {d\over dt} \delta_\alpha {\bf r}_j.
\end{equation}
This suffices to define the infinitesimal transformation of any function of particle positions, velocities, and time by transforming the argument of the function. In particular, the infinitesimal transformation of the Lagrangian (\ref{NewtLag}) satisfies
\begin{equation}
\delta_\alpha L = \frac{d}{dt} F_{\alpha},
\label{divNewtL}
\end{equation}
where
the non-vanishing components of $F_\alpha$ are given -- up to an additive constant -- by
\begin{equation}
F_7 =  MX, \quad F_8 =  M Y,\quad F_9 =  M Z,
\label{Fdef}
\end{equation}
and the Cartesian components of the center of mass ${\bf R}$ are denoted by $(X, Y, Z)$.

The commutators of  infinitesimal Galilean transformations applied to any function of particle positions, velocities, and time, {\it e.g.},  $L = L({\bf r}_1,\dots,{\bf r}_n, {\bf v}_1,\dots, {\bf v}_n, t)$, define the same Lie algebra as in eq.~(\ref{GalileiStructureEq}):
\begin{equation}
[\delta_\alpha, \delta_\beta] L= \sum_{\gamma=0}^9C_{\alpha\beta}^\gamma \delta_\gamma L,\quad \alpha, \beta, \gamma = 0,1,\dots, 9.
\label{prolongGalAl}
\end{equation} 
Using (\ref{divNewtL}) in (\ref{prolongGalAl}) yields
\begin{equation}
\frac{d}{dt}\left(\delta_\alpha F_\beta - \delta_\beta F_\alpha\right) =  \sum_{\gamma=0}^9C_{\alpha\beta}^\gamma \frac{dF_\gamma}{dt}.
\end{equation}
Therefore
\begin{equation}
\delta_\alpha F_\beta - \delta_\beta F_\alpha =  \sum_{\gamma=0}^9C_{\alpha\beta}^\gamma F_\gamma  + d_{\alpha\beta},
\label{Fdeq}
\end{equation}
where $d_{\alpha\beta}=-d_{\beta\alpha}$ are constants. For the Lagrangian (\ref{NewtLag}), where the functions $F_\alpha$ are defined by eq. (\ref{Fdef}), the non-zero values for these constants are
\begin{equation}
d_{17} = - d_{71} =  d_{28}=-d_{82} = d_{39} = - d_{93} = M.
\label{ddef}
\end{equation}

It follows from eq.~(\ref{Fdeq}) and the Jacobi identity that the constants $d_{\alpha\beta}$ satisfy the linear equations
\begin{equation}
\sum_{\delta=0}^9 \left(C^\delta_{\alpha\beta} d_{\gamma \delta} + C^\delta_{\beta\gamma} d_{\alpha \delta} +  C^\delta_{\gamma\alpha} d_{\beta \delta}\right) = 0.
\label{gclosed}
\end{equation}
This can be checked explicitly using the structure constants (\ref{structure}) and the definition of $d_{\alpha\beta}$ in eq.~(\ref{ddef}).
In the parlance of the theory of cohomology of Lie algebras, this last equation indicates that, for the Lie-algebra defined by $C_{\alpha\beta}^\gamma$, the 2-form defined by $d_{\alpha\beta}$ is {\it closed} using the natural exterior derivative operation defined for Lie algebras \cite{Azcarraga1995}.

With the above results concerning the Lie algebra of infinitesimal transformations in hand, we are ready to consider the question of the existence of a Lagrangian which is strictly invariant under the Galilei group. If there exists an equivalent Lagrangian $\tilde L$,
\begin{equation}
\tilde L = L - \frac{d}{dt} \Lambda,
\end{equation}
 that is invariant under the group, then the function $\Lambda$ must satisfy
\begin{equation}
\delta_\alpha \Lambda = F_\alpha + e_\alpha,
\label{deltaLambda}
\end{equation}
where $e_\alpha$ are constants.   The Frobenius  integrability conditions \cite{Schutz1980} for eq.~(\ref{deltaLambda}) can be calculated by applying $\delta_\beta$ to both sides of eq.~(\ref{deltaLambda}) and anti-symmetrizing with respect to $\alpha$ and $\beta$. The result is, after using eq.~(\ref{Fdeq}),\footnote{}
\begin{equation}
d_{\alpha\beta} =\sum_{\gamma=0}^9 C_{\alpha\beta}^\gamma  e_\gamma.
\label{gexact}
\end{equation} 
Recall that the functions $F_\alpha$ can be redefined by an additive  constant, but this just redefines the constants $e_\gamma$ in (\ref{deltaLambda}), (\ref{gexact}). Given $d_{\alpha\beta}$ (and $C_{\alpha\beta}^\gamma$), if there exists constants $e_\gamma$ such that eq.~(\ref{gexact}) holds then one can construct a group-invariant Lagrangian.\footnote{More precisely, one can find a Lagrangian invariant with respect to the transformations in the connected component of the identity transformation. This suffices to cover the boosts and translations.}  If no such constants exist, then there exists only a divergence-invariant Lagrangian. 
 
 In the language of Lie algebra cohomology, if there exists $e_\gamma$ such that $d_{\alpha\beta}$ can be expressed in the form shown in equation (\ref{gexact}) one says that the closed 2-form defined by $d_{\alpha\beta}$   is {\it exact}.  Moreover, it is straightforward to check that an array $d_{\alpha\beta}$ of the form shown in eq.~(\ref{gexact})  is automatically ``closed'', that is, it satisfies eq.~(\ref{gclosed}). Evidently, a divergence-invariant Lagrangian has an equivalent invariant Lagrangian if and only if the closed  2-form defined by $d_{\alpha\beta}$ is also exact.  More generally, given a symmetry group $G$, all divergence-invariant Lagrangians can be chosen to be strictly $G$-invariant if and only if all closed 2-forms on the Lie algebra of $G$ are exact \cite{Olver2022}.

The condition (\ref{gexact}) is a system of linear inhomogeneous equations for the $e_\gamma$.  Using the  constants given in equations (\ref{structure}) and (\ref{ddef})  it follows from a standard linear algebra analysis that there are no solutions $e_\gamma$  to the equations (\ref{gexact}), that is, the closed 2-form defined by eq.~(\ref{ddef}) is not exact.\footnote{Although we won't prove it here, it can be shown that the {\it only} closed 2-form which is not exact is, up to an overall  scaling, given by (\ref{ddef}).}
Therefore there are no equivalent Lagrangians which are strictly invariant under the entire Galilei group. 

If we restrict attention to the subgroup of the Galilei group consisting of space-time translations and orthogonal transformations, then it can be shown that all closed 2-forms are exact implying one can find an invariant Lagrangian for Newton's equations of motion.  Indeed, the Lagrangian (\ref{NewtLag}) is an example.  If we restrict attention to the subgroup of boosts, then again all closed 2-forms are exact and one can again find an invariant Lagrangian.  The Lagrangian (\ref{Ltilde}) is an example.  However, when one considers boosts and spatial translations together, there is no strictly invariant Lagrangian for Newtonian mechanics. 

\section{Central extensions for the infinitesimal Galilean algebra}
 \label{ProjSect}
 
Here we show how the same Lie algebra cohomology described in the previous section is responsible for the projective representation of the Galilei group.

Let us consider the infinitesimal form of the Galilean transformations of the wave function derived in \S4.   With respect to space-time translations and orthogonal transformations, the wave function behaves like a scalar, but for the boosts we must take account of the non-trivial transformation of the value of the wave function shown in eq.~(\ref{psitransboost}).   If ${\bf u} = \epsilon {\bf n}$ for a unit vector ${\bf n}$ and $\epsilon <<1$, we get
\begin{align}
\psi_{\epsilon{\bf n}}({\bf r}_1,\dots, {\bf r}_n, t) = &\psi({\bf r}_1,\dots, {\bf r}_n, t) + \epsilon\Big[ -t\sum_j {\bf n}\cdot \nabla_{{\bf r}_j} \psi({\bf r}_1,\dots, {\bf r}_n, t)\nonumber\\ 
&\quad \quad+ \frac{i}{\hbar}M {\bf n}\cdot {\bf R}\, \psi({\bf r}_1,\dots, {\bf r}_n, t)\Big] + {\cal O}(\epsilon^2),
\end{align}
so that the infinitesimal boosts transform the wave functions as follows:
\begin{equation}
\delta_7\psi = {\bf i} \cdot {\bf A},\quad \delta_8\psi = {\bf j}\cdot {\bf A},\quad \delta_9\psi = {\bf k}\cdot {\bf A},
\end{equation}
where
\begin{equation}
{\bf A} = -t\sum_j  \nabla_{{\bf r}_j} \psi({\bf r}_1,\dots, {\bf r}_n, t)
+ \frac{i}{\hbar}M {\bf R}\, \psi({\bf r}_1,\dots, {\bf r}_n, t).
\end{equation}
The first term in ${\bf A}$ comes from the infinitesimal transformation of the argument of the wave function.  The second term in ${\bf A}$ comes from the transformation of the complex value of the wave function.  Note that this latter infinitesimal transformation is a multiple of the identity transformation. 

Because of the multiples of the identity appearing in the representation of infinitesimal boosts of the wave function, the commutators of infinitesimal Galilean transformations of the wave function  yield a modified version of (\ref{GalileiStructureEq}):
\begin{equation}
[\delta_\alpha, \delta_\beta]\psi = \sum_{\gamma = 0}^{9}C_{\alpha\beta}^\gamma \delta_\gamma\psi + B_{\alpha\beta}\, \psi,
\label{GalPsi}
\end{equation}
where the non-vanishing components of the anti-symmetric array $B_{\alpha\beta} = -B_{\beta\alpha}$ are 
\begin{equation}
B_{17} = - B_{71} = B_{28} = - B_{82} = B_{39} = - B_{93} = \frac{i}{\hbar} M.
\label{GalCen}
\end{equation}

As an exercise,  the interested reader can repeat these calculations using the representation of the Galilei group described in \S5.  The definitions of the infinitesimal transformations will change, but the result is again given by (\ref{GalPsi}), (\ref{GalCen}).

The constants $B_{\alpha\beta}$ are proportional to the constants $d_{\alpha\beta}$ found in (\ref{ddef}).  Consequently they satisfy an identity of the form (\ref{gclosed}):
\begin{equation}
\sum_{\delta=0}^9 \left(C^\delta_{\alpha\beta} B_{\gamma \delta} + C^\delta_{\beta\gamma} B_{\alpha \delta} +  C^\delta_{\gamma\alpha} B_{\beta \delta}\right) = 0.
\label{centJ}
\end{equation}
 This identity is guaranteed by the Jacobi identity satisfied by the infinitesimal transformations and structure constants $C_{\alpha\beta}^\gamma$.
As before, the identity (\ref{centJ}) means that the  2-form on the Lie algebra defined by the constants $B_{\alpha\beta}$ is ``closed''.

The modification (\ref{GalPsi}) of the commutator algebra of infinitesimal Galilean transformations  is  called a ``central extension'' of that algebra.  The terminology arises because the identity transformation, which features in the second term on the right hand side of (\ref{GalPsi}), commutes with all the transformations $\delta_\alpha$ and thus represents a center of the algebra.  The appearance of this central extension is the infinitesimal version of the fact that the Galilei group is projectively represented in quantum mechanics.  In particular, from (\ref{GalCen}) we see that while the subgroup of the Galilei group consisting of space-time translations and orthogonal transformations has a bona fide representation  there is only a ``representation up to a phase'' for the subgroup consisting of spatial translations and boosts -- a result we encountered in \S\ref{QM2}.  For more on the central extensions and the projective representations of the Galilei group in quantum mechanics see \cite{Bargmann1954, Ballentine2014}.

 The algebra of infinitesimal Galilean transformations has been extended to include multiples of the identity, which reflects the quantum mechanical possibility of a representation up to a phase of the Galilei group.  Using multiples of the identity to redefine the infinitesimal transformations, the central extension can be rearranged to take  different forms.  Indeed, one can create a central extension of any Lie algebra of infinitesimal transformations, {\it e.g.},
 \begin{equation}
 [\widehat \delta_\alpha, \widehat \delta_\beta] = \sum_\gamma \widehat C_{\alpha\beta}^\gamma \widehat \delta_\gamma \,, 
 \end{equation}
 by simply shifting some of the infinitesimal transformations by a multiple of the identity (denoted {\bf 1}:)
 \begin{equation}
 \delta_\alpha^\prime = \widehat\delta_\alpha - \Gamma_\alpha {\bf 1}, \quad [ \delta_\alpha^\prime, \delta_\beta^\prime] = \sum_{\gamma}
 \widehat C_{\alpha\beta}^\gamma \delta_\gamma^\prime + Z_{\alpha\beta}{\bf 1},
 \end{equation} 
where
\begin{equation}
Z_{\alpha\beta} = \sum_\gamma \widehat C_{\alpha\beta}^\gamma \Gamma_\gamma .
\label{gex}
\end{equation}
This is the infinitesimal version of redefining the linear transformations representing a group with phase factors, thereby converting a representation into a projective representation. One is always free to adjust a central extension by such redefinitions.   In particular, if the central extension {\it only} involves terms of the form (\ref{gex}), then one can remove the central extension entirely by a simple redefinition of the transformations using multiples of the identity, namely,
\begin{equation}
\widehat \delta_\alpha =  \delta_\alpha^\prime + \Gamma_\alpha {\bf 1}.
\end{equation}   
This corresponds to redefining the linear transformations representing the group with phase factors such that a projective representation becomes a real representation. Notice that the structure (\ref{gex}) of $Z_{\alpha\beta}$ is the same as was found in (\ref{gexact});  such $Z_{\alpha\beta}$ are called ``exact'' 2-forms in the parlance of Lie algebra cohomology.

This naturally leads to the following question.  Is it possible to redefine the infinitesimal Galilean transformations using multiples of the identity to remove the central extension terms in eq.~(\ref{GalCen})?  We have again arrived at a question determined by the cohomology of the Lie algebra.  Specifically, all central extensions are associated to closed 2-forms, which is (\ref{centJ}), and they can be removed by redefining the infinitesimal generators if the 2-forms are also exact, that is, there exist constants $E_\alpha$ such that
\begin{equation}
B_{\alpha\beta} = \sum_{\gamma = 0}^{9}  C_{\alpha\beta}^\gamma E_\gamma .
\end{equation}  
The  anti-symmetric array $B_{\alpha\beta}$ is, up to an overall scaling,
precisely the same as appears in  (\ref{ddef}), which we saw is not exact. Thus the central extension appearing in (\ref{GalPsi}) cannot be removed by adding multiples of the identity to the generators.  Moreover, this fact is established using the same result which establishes the non-existence of a Galilean-invariant Lagrangian for Newton's equations of motion!   As we have seen, this  is not a coincidence but can be understood from  Feynman's path integral formulation of quantum mechanics.

\end{document}